\documentclass[conference]{IEEEtran}
\usepackage[T1]{fontenc}
\usepackage[utf8]{inputenc} 
\usepackage{flushend}
\usepackage{hyperref}
\usepackage{fixltx2e}
\usepackage{cite}

\begin{document}

%
\title{The Affect of Software Developers: Common Misconceptions and Measurements}

\author{
\IEEEauthorblockN{Daniel Graziotin, Xiaofeng Wang, Pekka Abrahamsson}
\IEEEauthorblockA{Faculty of Computer Science\\
Free University of Bozen-Bolzano\\
Bolzano/Bozen, Italy\\
Email: \{daniel.graziotin, xiaofeng.wang, pekka.abrahamsson\}@unibz.it}
}

\maketitle

\begin{abstract}
The study of affects (i.e., emotions, moods) in the workplace has received a lot of attention in the last 15 years. Despite the fact that software development has been shown to be intellectual, creative, and driven by cognitive activities, and that affects have a deep influence on cognitive activities, software engineering research lacks an understanding of the affects of software developers. This note provides (1) common misconceptions of affects when dealing with job satisfaction, motivation, commitment, well-being, and happiness; (2) validated measurement instruments for affect measurement; and (3) our recommendations when measuring the affects of software developers.
\end{abstract}


\IEEEpeerreviewmaketitle

\section{Introduction}\label{introduction}

Affects---emotions and moods---play a role in people's daily job; they
pervade organizations, the relationships between workers, deadlines,
work motivation, sense-making, and human-resource processes \cite{Barsade2007}.
Although emotions have been historically neglected in studies of
industrial and organizational psychology, an interest in the role of
affects on job outcomes has accelerated over the past fifteen years in
organizational psychology \cite{Fisher2000c}, and lately, in software engineering \cite{Novielli2014}. Still, software engineering research is lacking an
understanding of how affects have a role in the software construction
process \cite{Khan2010}. This paper builds upon our experience on the affects of software developers
\footnote{
\url{http://dx.doi.org/10.7717/peerj.289},
\url{http://dx.doi.org/10.1002/smr.1673},
\url{http://dx.doi.org/10.1007/978-3-642-39259-7\_7}, and
\url{http://dx.doi.org/10.1109/MS.2014.94}.
See the first three for definitions. See the last one for a literature review of the software engineering studies.}.
It reports (1) common misconceptions of affects
when job satisfaction, motivation, commitment, well-being, and happiness
are dealt with in literature; (2) validated measurement instruments for
affect measurement; and (3) our recommendations when measuring the
affects of software developers.

\section{Common Misconceptions}\label{common-misconceptions}

We have found that there are some common misconceptions in literature
regarding affects. The first is \emph{job satisfaction}, which is often
confused with affects in a workplace. Job satisfaction is an attitude,
not an affect \cite{Brief1998b}. An attitude is an evaluative judgment made with
regard to an attitudinal object, in this case one's job \cite{Weiss2002}. The
current definitions of job satisfaction ``\emph{have obscured the
differences among three related but distinct constructs: evaluations of
jobs, beliefs about jobs, and affective experiences on jobs}.''
(\cite{Weiss2002}, p. 173). More precisely, job satisfaction is ``\emph{a
positive (or negative) evaluative judgment one makes about one's job or
job situation}.'' (\cite{Weiss2002}, p. 175).

Affects are not \emph{motivation}, either. Mitchell defined motivation
as ``\emph{those psychological processes that cause the arousal,
direction, and persistence of voluntary actions that are goal
directed.}'' (\cite{Mitchell1982}, p. 81). This definition already suggests that
motivation is not an affect, however the two constructs appear to be
related. According to Seifert {\cite{Seifert2004}, when presented with a task,
individuals perform evaluative judgments about the task itself, and they
respond affectively based upon task and personal characteristics. These
generated affects dictate successive motivation towards the task.

\emph{Commitment} has been defined as a psychological state of
attachment that defines the relationship between a person and an entity
(organization) \cite{OReilly1986}. Commitment is multifaceted. According to Meyer
and Allen \cite{Meyer1991}, commitment is conceptualized in the forms of
affective, normative, and continuance commitment. While normative and
continuance commitment deal with perceived moral obligations and the
awareness of the costs associated with leaving the organization
respectively, affective commitment refers to an employee's attachment
to, identification with, and involvement within an entity, e.g., an
organization, a project, or a team.

Similar to job satisfaction, \emph{well-being} is an attitude \cite{Diener1984}.
Subjective well-being consists of two interrelated components, which are
life satisfaction that refers to a cognitive sense of satisfaction with
life \cite{Diener1984}, and positive and negative affects. Subjective well-being
can be considered as one's self-evaluation of life, which is influenced
by affects.

\emph{Happiness}, on the other hand, is a complex construct, which has
different psychological and philosophical definitions. We adhere to
Haybron's \cite{Haybron2005} view of happiness as a matter of a person's affective condition, where only central affects are concerned. An individual is
happy if the person's affect balance is characterized by frequent
central positive affects. As a counter-example, the pleasure of eating a
cracker is not enduring and probably not affecting happiness; therefore,
it is considered a peripheral affect.

The affects of individuals are related to all the above-mentioned
constructs. Affective reactions of the individuals influence their job
satisfaction \cite{Hume2008}, motivation \cite{Mitchell1982}, affective commitment
\cite{Rhoades2001}, well-being \cite{Diener1984}, and happiness \cite{Haybron2005}.

\section{Measuring Affects and
Recommendations}\label{measuring-affects-and-recommendations}

We recommend employing the Scale of Positive and Negative Experience
(SPANE) \cite{Diener2009} questionnaire for assessing the affects of software
developers when it is not necessary to understand the affects raised by
a particular stimulus. SPANE assesses a broad range of pleasant and
unpleasant affects by asking the participants to report them in terms of
their frequencies during the last four weeks. It is a 12-item scale,
divided into two sub-scales of positive affects and negative affects.
The answers to the items are given on a five-point scale ranging from 1
(very rarely or never) to 5 (very often or always). The aggregated
scores result in the Affect Balance Score (SPANE-B). SPANE-B is an
indicator, with range {[}-24; +24{]} of how \emph{happy} or
\emph{unhappy} people are in terms of how often they feel positive and
negative affects. SPANE has been validated to provide good psychometric
properties and to converge to other measurement instruments of affects
\cite{Diener2009}.

Instead, for assessing the affects of software developers triggered by a
stimulus, e.g., a development tool, a user interface, or a development
task, the Self-Assessment Manikin (SAM) \cite{Lang1994} is recommended. SAM is
a pictorial, i.e., non-verbal, assessment method. SAM produces three
measures for understanding aggregated affects, namely valence, arousal,
and dominance associated with a person's affective reaction to an object
(or a stimulus) \cite{Lang1994}. For example, for a five-point rating scale, a
value of five for valence means ``very high attractiveness and pleasure
towards the stimulus."

Many psychological measures, including those reported here, need special
attention on the within- and between-subjects analyses of repeated
measurements. First, the assessed metrics are not transferrable or
stable across persons. For example, an assessed \emph{one} for an
individual's arousal may be equal to a \emph{three} for a different
individual. However, ``\emph{it is sensible to assume that there is a
reasonable stable metric within persons}'' (\cite{Hektner2007}, p. 10). For having
comparable measurements between subjects, the raw scores are transformed
into standard scores (also known as z-scores). The measurements become
``dimensionless'', thus comparable with those of other participants
\cite{Hektner2007}. Second, the repeated measurements often present two layers of dependency of data: at the \emph{participant} level and the \emph{time}
level, grouped by the participant. While the analysis of variance family
provides \emph{rANOVA} as a variant for repeated measurements,
\emph{ANOVA} procedures are discouraged {\cite{Gueorguieva2004} in favor of
\emph{linear} \emph{mixed-effects} (\emph{LME}) models.

Our last note is on the misuse of the term \emph{psychometrics in}
previous software engineering research. So far psychometrics has been
employed to mean \emph{psychological measurements} \cite{Feldt2008}. However,
\emph{psychometrics} is the field of study concerned with the
implementation and validation of psychological measurements. A
measurement instrument in psychology has to possess acceptable validity
and reliability properties, which are provided in psychometric studies
of the measurement instrument. A modification to an existing measurement
instrument (e.g., adding, deleting, or rewording items) often requires a
new psychometric study. For similar reasons and for ensuring higher
reproducibility, the participant's instructions should be made available
with a paper, because the instructions might influence the participants'
responses.



%

\bibliographystyle{IEEEtran}
\bibliography{References}


\newpage

\onecolumn

\noindent
This is the preprint of the following article:

\vspace{10 mm}

\fbox{
    \parbox{\textwidth}{
    	Citation: Graziotin, D., Wang, X., \& Abrahamsson, P. (2015). The Affect of Software Developers: Common Misconceptions and Measurements. In 2015 IEEE/ACM 8th International Workshop on Cooperative and Human Aspects of Software Engineering (CHASE 2015) (pp. 123–124). Firenze, Italy: IEEE Computer Society. DOI: \href{10.1109/CHASE.2015.23}{http://dx.doi.org/10.1109/CHASE.2015.23}
    }
}

\vspace{10 mm}

\textcopyright IEEE. Personal use of this material is permitted. However, permission to reprint/republish this material for advertising or promotional purposes or for creating new collective works for resale or redistribution to servers or lists, or to reuse any copyrighted component of this work in other works must be obtained from the IEEE.

This material is presented to ensure timely dissemination of scholarly and technical work. Copyright and all rights therein are retained by authors or by other copyright holders. All persons copying this information are expected to adhere to the terms and constraints invoked by each author's copyright. In most cases, these works may not be reposted without the explicit permission of the copyright holder.

For more details, see the IEEE Copyright Policy.

\end{document}